\documentclass[doublecol]{epl2} 
\usepackage{graphicx}
\usepackage{color}
\usepackage{multirow}
\usepackage{hhline}
\usepackage[normalem]{ulem}

\title{Is your EPL attractive? Classification of publications through download {statistics}}
\shorttitle{Is your EPL attractive?} 

\author{O.~Mryglod\inst{1} \and R.~Kenna\inst{2} \and Yu.~Holovatch\inst{1}}
\shortauthor{O.~Mryglod \etal}

\institute{
  \inst{1} Institute for Condensed Matter Physics of the National Acad. Sci. of Ukraine, \\
             79011 Lviv, Ukraine\\
  \inst{2} Applied Mathematics Research Centre, Coventry University,
              Coventry, CV1 5FB, England
}
\pacs{01.75.+m}{Science and society }
\pacs{89.65.-s}{Social and economic systems}
\pacs{89.75.-k}{Complex systems}

\abstract{Here we consider {the download statistics of \emph{EPL} publications. We find that papers in the journal are} characterised by fast accumulation{s} of downloads during the first couple of months after publication, followed by {slower rates} thereafter{, behaviour which} can be {represented by a model} {with predictive power}. We also find that individual papers can be classified in various ways{, allowing us to compare categories for open-access and non-open-access papers}.
{For example, for the latter publications, which comprise the bulk of \emph{EPL} papers,} a small proportion (2\%) display intense bursts of download activity, possibly  following an extended period of {less} remarkable behaviour. About {18\%} have an especially high degree of attractiveness {over and above what is typical for the journal}.
One can also classify the ageing of attractiveness by examining download half-lives.
Approximately {18\%} have strong interest initially, waning in time. A further {20\%} exhibit ``delayed recognition'' with relatively late spurs in download activity. {Although open-access papers enjoy more downloads on average, the proportions falling into each category are similar. }
}
\begin{document}

\maketitle

\section{Introduction}

The dissemination of scholarly research through publication in peer-reviewed journals lies at the heart of the academic process.
Having one's research peer reviewed and published is, however, only the start of the journey to academic recognition, even with journals of high repute.
Increasingly, citation-based metrics have come to be used as indicators of the value
of individual research papers and, indeed, of the calibre of their authors. Journals are also interested in measurements of performance and the use of {\emph{impact factors}} is now ubiquitous.
Citation-based estimates of individual papers, researchers and journals have become so important to  scientists that, rather than leaving it entirely in the hands of scientometricians, managers and policy makers, many physicists have taken to investigate various aspects of measuring science themselves \cite{Schreiber1,Schreiber2,Alfi1,Alfi2,MHM,Acuna,Thurner2011,us1,us2}.
Besides these  practical motivations for physicists to turn the scientific process on itself, there are also curiosity drivers, and performance data associated with scientific publications are amenable to physicists' tools \cite{Alfi1,Alfi2}.
{M}otivated by both practical considerations and by curiosity, then, we sought to investigate the download {statistics} of {\emph{EPL}}.
We targetted {\emph{EPL}} as ``Europe's flagship letters journal'' of broad interest to the physics community.
An understanding of the downloading process here may deliver a viewpoint on the value of scientific outputs {complementary to others in current use}, in particular their attractiveness, and may inspire similar studies for other journals, leading to new ways to compare them.

\section{Peer review, downloads and citations}

The peer-review process is generally recognised by academics as the most reliable basis for evaluation of research \cite{Wasserman}.
While the strengths associated with expert judgement are documented, the process is not perfect and drawbacks include subjectivity of referees, bias and the fact that the process is  labour intensive and time consuming \cite{Thurner2011,Wasserman,Bellis2009,Bornmann2012}.

While citations are considered as indicators of {academic impact}, it is widely recognised that  impact and research quality are not the same \cite{us1,us2}.
The imperfections of citation counting are also well documented and include gratuitous citing;  bias; citation inflation; retrieval issues; discipline traditions, and the fact that many cited papers are not read by citing authors {\cite{Simkin,Wesel2014}}.
The long time lags which are necessary to accumulate numbers of citations represent another drawback. Despite all of these issues and more, the popularity of citation-based metrics amongst decision makers persists, because they deliver automatic, rapid and cheap information. These {drawbacks} have driven the search for alternative and complementary approaches.

\looseness=-1 {There} is a weak but positive correlation between  citation and download numbers
and undoubtedly some of the factors which limit other metrics as a measure also apply to downloads \cite{Perneger2004,Watson2009,Kurtz2004,Moed2005,Bollen2005,Brody2005}.
Therefore, while downloads cannot be considered as a substitute for either citation-based or peer-review-based approaches, they can  play a complementary role.
While peer review and citations reflect opinion about a paper's quality and scientific impact after reading, downloads rather reflect interest before{hand} \cite{Watson2009}.
In other words, in addition to popularity and prestige  \cite{Franceschet2010},  papers may be distinguished by their \emph{attractiveness}. In such a classification, the overall number of citations measures  popularity, the number of important citations {\cite{Franceschet2010}} is evidence of prestige, whereas the number of downloads reflects the level of attractiveness of a publication.
Here we wish to analyse download {statistics} {to study attractiveness of {\emph{EPL}} papers.}

\section{Download data}

{\emph{IOPscience}} is the online service for the journal{s} of the {\emph{Institute of Physics (IoP)}}.
We were provided with data on the full-text downloads for papers, published  in {\emph{EPL}} between January 2007  to June 2013 with one month resolution.
The downloads are counted on an IP-address basis with multiple requests made from the same address considered as separate downloads. Only full-text downloads {from} the {\emph{IOPscience}} web-pages are counted.
The data are automatically cleansed of suspicious and robot activity, and are  COUNTER compliant.
 {For the purposes of our analysis, papers published online after the 25$^{\mathrm{th}}$ day of a given month were allocated to the following month because of the small lag time for download accumulations. All {\emph{EPL}} papers are freely downloadable for 2 months after online publication.
The open access {(OA)} period can be extended in some cases and such papers  have  different download statistics as evidenced in  Fig.~\ref{Fig_sync_Moed}.}
{There} is an increasing number of ways to download {\emph{EPL}} articles without going through {\emph{IOPscience}}, most notably from the \emph{arXiv} and personal web-sites.
In 2013, in particular, counts to all IOP hosted journals were down by about 10\% on 2012  probably due to {\emph{Google}} modifying their search algorithm in favour of
articles that are not behind paywalls.

\looseness=-1With these various aspects and caveats in mind, we proceed to report on our case study of the analysis of publications downloads and aging in {\emph{EPL}}.
{Our data set comprises 4\,986 non-OA papers and 377 OA papers. }
We use both synchronous and diachronous approaches \cite{Moed2005}.
The synchronous approach is calendar based, wherein Month 1 refers to January 2007, Month 2 to February 2007, and so on, with Month 78 referring to June 2013.
The synchronous approach is appropriate to describe the downloading patterns of the entire journal as opposed to individual papers.
The diachronous considers time since an individual paper's publication.
In this case, Month $n$ is article-specific and refers to the $n$th month after the paper's publication online.
The diachronous approach therefore corresponds to download patterns for individual papers.

\section{Synchronous approach}
 \begin{figure}[!t]
\centerline{\includegraphics[width=0.43\textwidth]{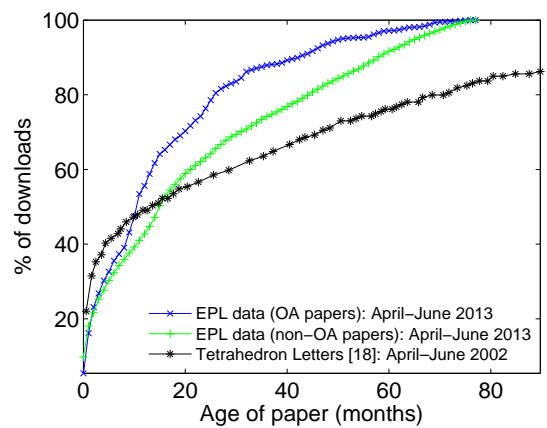}}%
\caption{(Color online)
Relative cumulative distributions of ages of downloaded papers.
{The `+' symbols, (green) represent non-OA {\emph{EPL}} papers,
`$\times$' (blue) represent OA {\emph{EPL}} papers} and `{\large{$\ast$}}' (black)
represent data from {\emph{Tetrahedron Letters}}~\cite{Moed2005}.}
\label{Fig_sync_Moed}
\end{figure}
\looseness=-1Since January 2007, an average of 64 papers have been published online in {\emph{EPL}} each month.
Following ref.~\cite{Moed2005}, one can construct cumulative distribution curves by age, starting on given dates.
The digitised version of Fig.~1 from ref.~\cite{Moed2005} which corresponds to download data for the journal {\emph{Tetrahedron Letters}} \cite{TetrahedronLett} is presented in Fig.~\ref{Fig_sync_Moed} along with the corresponding {\emph{EPL}} data {both for OA and non-OA papers} downloaded during April--June 2013.
Data labeled $(x,y)$ indicate that {within given period} $y\%$ of downloads were to papers $x$ months old or less.
In ref.~\cite{Moed2005} a critical point  of 3 months was identified, as a demarcation between two different regimes characterised by different slopes to the left and right of the \emph{Tetrahedron Letters} curve (see also \cite{Watson2009}). The difference was ascribed to  recent publications being downloaded because of their novelty and older papers for archiving, background reading or similar.
The slope{s} corresponding to the {\emph{EPL}} curve{s do} not change so dramatically, although softer transition{s are} evident.
One may conjecture that the difference between the two {journals}  could be caused by disciplinary peculiarities or different rules for access.
(\emph{Tetrahedron Letters} does not have an initial  2-month period of open-accessibility comparable  to that of \emph{EPL}.)

\begin{figure}[t]
\centerline{\includegraphics[width=0.43\textwidth]{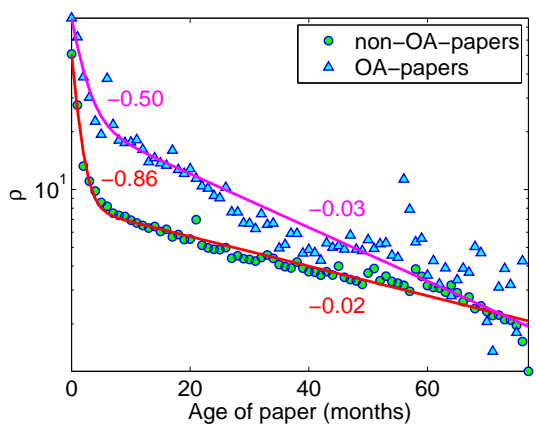}}
\caption{(Color online) {Density of downloads per paper ($\rho$) versus papers' ages}.
{The solid curves show} the predictions of the model (\ref{eq_our_model}) and the corresponding exponential decay constants are indicated.}
\label{Fig_syncr_means_decay}
\end{figure}

\looseness=-1 To understand the shape of the cumulative distributions depicted in Fig.~\ref{Fig_sync_Moed}, we consider the non-cumulative data in Fig.~\ref{Fig_syncr_means_decay}.
Here we plot {the density of downloads $\rho$ -- defined as  mean numbers of downloads per paper -- against their age}.
Two different regimes are again distinguished;
new papers attract more intense activity  while after 5 or 6 months the download speed decreases. One may attribute the enhancing of download frequencies during the first few months to the novelty and free accessibility.
%

In order to fit to the two regions of Fig.~\ref{Fig_syncr_means_decay}, it is convenient to use {a} \emph{two-factor model} \cite{Moed2005}:
{
\begin{eqnarray}
\label{eq_our_model}
\rho(t)=\rho_0\left[ A\exp(-b_1 t)+ (1-A)\exp(-b_2 t)\right], \\
\nonumber 0\leq A\leq1, \qquad b_1>0, \qquad b_2>0\,,
\end{eqnarray}}
where $A$ and $(1-A)$ are relative weights of the two factors (two different motives for downloads) and {$\rho_0$ is the density of downloads which corresponds to the newest papers (published in the month of downloading)}.
The parameters $b_1$ and $b_2$ are    exponential decay constants
corresponding to early and later download patterns.
Using  nonlinear-curve least-squares fitting, we obtain the estimates {for non-OA papers $A \approx 0.84$, $b_1 \approx 0.86$, $b_2 \approx 0.02$} {(parameters close to the estimates of \cite{Moed2005} for {\emph{Tetrahedron Letters}}: $A \approx 0.92$, $b_1 \approx 0.50$, $b_2 \approx 0.014$)}. {The corresponding estimations for OA papers are  {$A \approx  0.71$, $b_1 \approx 0.50$, $b_2 \approx 0.03$}. Thus, the two regimes are observable for both data sets, but the slopes are different. The OA downloads are more concentrated on the first months after publication online.}

\begin{figure}[t]
\includegraphics[width=0.43\textwidth]{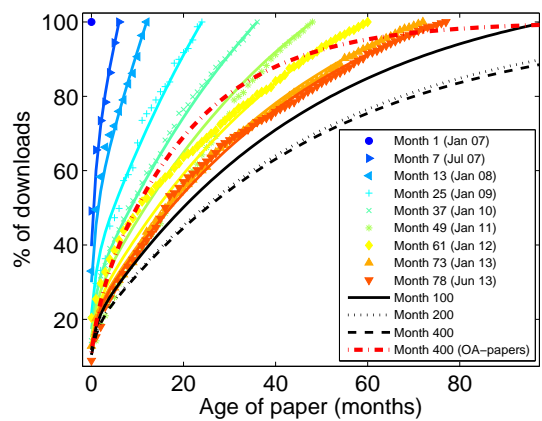}%
\caption{%
(Color online) Cumulative distributions of ages of papers downloaded during different months for {\emph{EPL}}. {The  model (\ref{eq_our_model}) delivers the curves  and predictions for long-term behaviour.}
}%
\label{Fig_sync}%
\end{figure}

Fig.~\ref{Fig_sync} is similar to Fig.~\ref{Fig_sync_Moed} except (a) all the data refer to {\emph{EPL}} and (b) the download windows (each of one month's duration) vary, each period corresponding to a different colour {(online)}.
Since no particular seasonal effects were observed, of the 78 months for which we have data, only 10 representative months are depicted for the plot, for clarity.
These are Month 1 (January 2007), Month 7 (July 2007) \dots Month 78 (June 2013) as listed in the figure. The cumulative curves, based on {a two-factor} model are indicated by solid lines.
Downloads to papers during Month 1 (January 2007) can, of course, only be to papers which have appeared during that month, so the only data point has $y=100\%$. At the other extreme, downloads during June 2013 can be to papers up to $x=77$ months old.

The plot in Fig.~\ref{Fig_sync} can be used to determine ages of the papers required to provide a certain percentage of downloads within a particular month.
For example,  $y=50\%$ of {non-OA}  downloads during January 2013 were to papers $x=17$ months old or less.

Our model also allows us to see the approach of the cumulative curves to a limiting distribution {and} the predictions of the model using extrapolated values for the $100^\mathrm{th}, 200^\mathrm{th}$ and $ 400^\mathrm{th}$ months are presented for non-OA papers.  {These can be compared to the predictions for OA papers, which are also shown in the figure.}

{This  allows us to make   predictions for the long-term behaviour.
For example, in the long term, typically $y=50\%$ of non-OA downloads during a given month are to papers $x=25$  months old or less.
For OA papers, this proportion of downloads are typically to papers $x=10$ months old or less.
Thus OA papers accumulate given proportions of their downloads faster than non-OA papers. }

\section{Diachronous approach}
\begin{figure}[t]
\centerline{\includegraphics[width=0.43\textwidth]{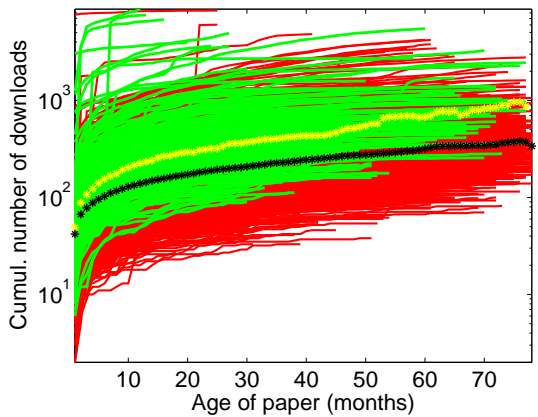}}%
\caption{%
(Color online) Cumulative number of downloads for each paper separately{: dark lines (red) represent non-OA-papers while light lines (green) represent OA-papers}. The median values for papers of the same age are indicated by the symbols $\large{\ast}$ {(black for non-OA-papers and yellow for OA-papers)}.}%
\label{Fig_separately_all}%
\end{figure}

Cumulative downloads of individual papers are plotted in Fig.~\ref{Fig_separately_all}. The plot indicates the diversity of individual download histories, the median values of which can be considered as typical.
This typical behaviour is fast accumulation of downloads during the first couple of months, followed by a slower rate  thereafter.
{The OA papers tend to accumulated downloads more rapidly than non-OA papers. Correspondingly, the median numbers of downloads for such papers are  consistently higher.}
We are also interested in unusual patterns of downloads and our next concern is how to detect  papers which {exhibit} such behaviour.
%
%

\subsection{``Bursty'' papers} 
\begin{figure}[t]
\centerline{\includegraphics[width=0.43\textwidth]{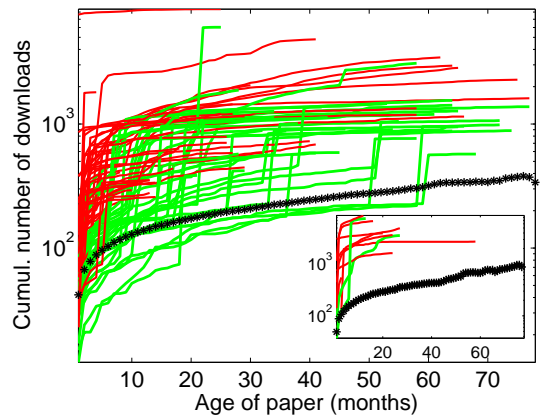}}%
\caption{%
(Color online) Cumulative numbers of downloads for bursty papers.
{The main panel is for non-OA paper and the equivalent plot for OA papers is shown in the inset.}
The median values  are represented by the symbols $\large{\ast}$ (in black).
`Sleeping beauties' are highlighted (in green online).
}%
\label{Fig_separately_exotic}%
\end{figure}

A simple but straightforward way to detect bursts in download activity is to flag strong deviations from an individual paper's typical download pattern.
Let $\sigma_i{(T)}$ be the standard deviation in the number of downloads of paper $i$ over its entire history {to month $T$}.
Let $\sigma_{\mathrm{med}}(T)$ be the standard deviation of {the} median values  over the first $T$ months.
Define
\begin{equation}
\Delta_i(T)=| \sigma_i{(T)}- \sigma_{\mathrm{med}}(T) |.
\end{equation}
Each paper is characterized by a value of $\Delta_i(T)$.
Bursts in activity may then be flagged by large values of $\Delta_i(T)$ compared to the average value $\langle \Delta_i(T)\rangle$. Of the 4\,986 {non-OA} papers analysed, three have {$\Delta_i(T) /\langle \Delta_i(T)\rangle \geq 100$} for certain values of $T$.
Thirty-seven papers have {$\Delta_i(T) / \langle \Delta_i(T)\rangle \geq 10 $} and  ninety-six have  {$\Delta_i(T)/ \langle \Delta_i(T)\rangle \geq  5$}.
Since this represents a suitably small fraction of all papers ($\approx 2\%$), we choose
 $ \Delta_i(T)/ \langle \Delta_i(T)\rangle = 5$ as a critical value.
The download histories of each of these 96 bursty papers {are} depicted in Fig.~\ref{Fig_separately_exotic}, alongside the  median pattern (for all papers). {Ten OA papers (2.7\% of all OA papers) can be described as bursty using the criteria described above and are shown in the inset of Fig.~\ref{Fig_separately_exotic}.}

Some of the bursty papers exhibit the phenomenon of delayed recognition. Their download patterns are unremarkable for an extended period after which a burst of activity occurs. We describe such papers as ``sleeping beauties'' \cite{SB1,SB2,SB3}.
{Thirty-five} {non-OA} sleeping beauties {(two  OA papers)}, for which the download burst{s occur later than} 6 months after publication, are {highlighted} in Fig.~\ref{Fig_separately_exotic}.
The bursts in activity for the remaining {bursty} papers come sooner after publication.
{For a summary of these, and other, results, see Table~1.}
To continue our analysis of usual {\emph{EPL}} patterns, the bursty  papers are excluded. Therefore we continue with 4\,890 {non-OA} papers {and 367 OA papers}.

\subsection{Diversity of download patterns}
\begin{figure}[t]
\centerline{\includegraphics[width=0.43\textwidth]{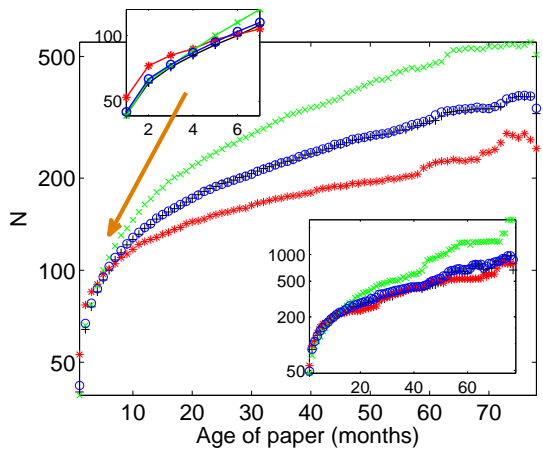}}%
\caption{%
(Color online) {Median values for cumulative numbers of downloads categorised according to half-lives:  `+' symbols (black) represent {`usual'} download patterns,
`{\large{$\ast$}}' (red) are flashes-in-the-pan, and `$\times$' (green) represent delayed downloads.
The corresponding values for all non-bursty papers are presented by blue circles. The main panel is for non-OA papers with  the upper inset a blow-up for young papers. {The same plot for OA papers is shown in the lower inset.}}} %
\label{Fig_diff_medians}%
\end{figure}

To categorise download patterns, and thereby the ageing of the attractiveness of {\em{EPL}}  papers, we borrow the notion of {\emph{durability}} which was developed for citation histories in ref.~\cite{Costas2010}.
The cornerstone of this approach is to consider  download half-lives.
The half-life $M^{50}(t)$ is the number of months by which a paper achieves 50\% of its current downloads.
The statistics of these values allow us to find  papers with  {usual}, fast and slow download rates.
We consider as  {`usual'} the 3\,065 {non-OA} papers for which $M_i^{50}$ is between the 25th and 75th percentiles (P25 and P75). These represent {62\%} of the 4\,890 {non-OA papers} considered.
If $M_i^{50}$ is smaller than {the} P25 value it means the downloads are initially accrued faster than  {`usual'} and tail off later.
We refer to  the 869 papers (18\%) with this pattern as `flashes in the pan'.
If $M_i^{50}$ is larger than P75 value, the paper has begun its life with a download rate {lower than average}, only to accelerate later. The 956 such papers (20\%) are termed  `delayed'. {The corresponding proportions for OA papers are similar: $ 65\%$ have usual ageing behaviour and there are $ 17.5\%$ of the papers in each of the others two categories.
However, while `flashes-in-the-pan' and `delayed' non-OA papers settle into their final ageing regimes already 4 or 5 months after publication (see the upper inset in Fig.~\ref{Fig_diff_medians}), for OA papers it takes a longer time period of about $13$ months. }

Median download {statistics} for {the above defined} three categories are depicted in Fig.~\ref{Fig_diff_medians}.
The aging of downloads for  {`usual'}  papers  goes more or less in a same way as the overall median pattern.
Flashes-in-the-pan papers rapidly collect downloads after online publication but later the rate of downloads decreases. It is noticeable that the median values for such papers in the long term tend to be lower than the median values for {`usual'} papers.
Delayed papers are characterized by a slightly slower downloads at very beginning but they are well-downloaded later.
(Sometimes, the term ``delayed recognition'' has been used synonymously with ``sleeping beauties'' but we  distinguish between these two terms in order to differentiate between the papers with unexpectedly large bursts of downloads -- sleeping beauties -- and those characterized by gradual changing of download accumulation  -- delayed).

%
\begin{figure}[t]
\centerline{\includegraphics[width=0.445\textwidth]{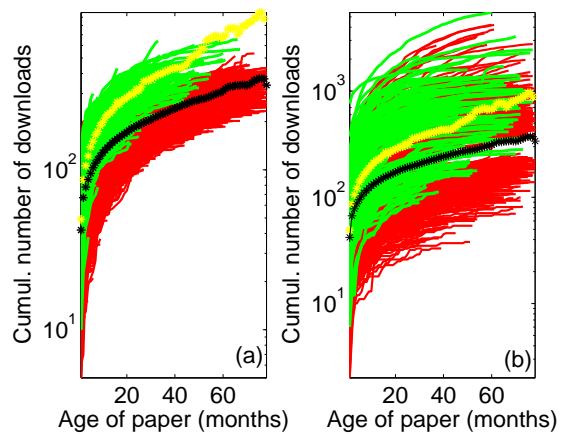}}%
\caption{%
(Color online) Cumulative number of downloads for (a) typical and (b) atypical papers {separately: dark (red) lines represent non-OA papers while the light (green) ones represent OA papers.
 The median values for papers of the same age are indicated by the symbols  $\large{\ast}$ (black for non-OA papers and yellow for OA papers)}. }%
\label{Fig_separately_typical-nontypical}%
\end{figure}

Finally we wish to address the question of overall attractiveness of papers, as measured by downloads.
The root-mean-square deviation (RMSD) of cumulative numbers of downloads from the standard values can be used to gauge the extent to which  patterns of downloads for individual papers differ from the standard one based on medians.
The value  $\mathrm{RMSD}_\mathrm{c}=66$ was taken as critical\footnote{{This choice was made because of an empirical observation that the number of papers with $\mathrm{RMSD}$ above criticality decays slowly for $\mathrm{RMSD}_\mathrm{c}<66$ and rapidly for $\mathrm{RMSD}_\mathrm{c}>66$}.} to filter the {non-OA} papers close to the standard.
In this way, 2\,926 ({60\%}) `typical' papers were identified ($\mathrm{RMSD}<\mathrm{RMSD}_\mathrm{c}$), with the remaining   1\,964 (40\%) as `atypical', see Fig.~\ref{Fig_separately_typical-nontypical}.
{Using the critical value $\mathrm{RMSD}_\mathrm{c}=105$ for OA papers, 185 ($50\%$) `typical' and 182 (the other $50\%$) `atypical' papers were classified (see Fig.~\ref{Fig_separately_typical-nontypical}). For both categories OA papers accumulate constantly more downloads on average comparing to non-OA papers.}

\begin{figure}[t]
\centerline{\includegraphics[width=0.43\textwidth]{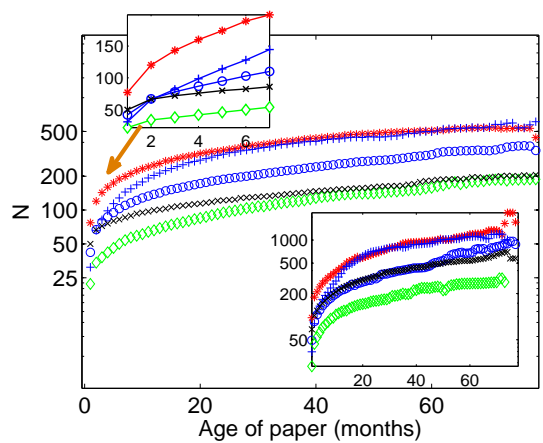}}%
\caption{%
(Color online) Median values for cumulative number of downloads for {typical} as well as for different categories of {atypical} papers:
highly downloaded (red `{\large{$\ast$}}' symbols),
weakly downloaded (green `{\large{$\diamond$}}'),
weakly-then-strongly downloaded (bl{ue} `+')
and strongly-then-weakly downloaded (black `$\times$'). {The corresponding values for all non-bursty papers are presented by blue circles. }
The {upper} inset is a blow-up for young papers. {The same plot for OA papers is shown in the lower inset.}
}%
\label{Fig_diff_medians2}%
\end{figure}

The atypical papers {(see Fig.~\ref{Fig_separately_typical-nontypical} (b))} are analysed to see the most notable examples of highly or poorly downloaded papers. {Among the non-OA papers t}here are 857 atypical {ones} ({18\%} of all papers), which are characterized by persistently higher cumulative download values compared to the standard.
{An}other 581 atypical papers (12\% of all papers) are always less downloaded, see Fig.~\ref{Fig_diff_medians2}. The cumulative curves of downloads for the remaining  526  atypical papers (10\%) cross the standard curve {from below (6\%) or from above (4\%)} similarly to the flashes-in-the-pans and delayed (see Fig.~\ref{Fig_diff_medians} and Fig.~\ref{Fig_diff_medians2}). {The proportion of OA papers for all these categories is again quite similar, see Table~1: $22\%$ of atypical papers are more attractive than average, $15\%$ are less attractive and for $13\%$ the attractiveness is changing with time. }
Here, as in Fig.~\ref{Fig_diff_medians}, one observes that the ageing behaviours  for some of the categories look similar (see the lower inset in Fig.~\ref{Fig_diff_medians2}).

\section{Perspectives and discussion}
\begin{table}[!b]
\vspace{-5mm}
\label{tab}
\caption{Categorisation of {\em{EPL}}  papers according to downloads {(the data for OA papers are in brackets)}.
}
\vspace{2ex}
\footnotesize{
\begin{tabular}{|l|l|l|}
\hline
\multicolumn{3}{|c|}{\parbox[t]{8cm}{\textbf{Categorisation by burstiness}}}\\
\hline
\multirow{3}{*}{\parbox[t]{1.5cm}{4\,986 {(377)} papers }} &\multicolumn{2}{|l|}{ 98\% are ``non-bursty'' papers }\\
 \hhline{~--}
&\multirow{3}{*}{\parbox[t]{1.6cm}{2\% {(3\%)} are ``bursty'' papers}}&\parbox[t]{4.45cm}{{1\% (1\%)} are ``sleeping beauties''}\\
& & {1\% (2\%)} burst early  \\
\hline
\hline
\multicolumn{3}{|c|}{\parbox[t]{8cm}{\textbf{Categorisation by half-lives (ageing of attractiveness)}}}\\
\hline
\multirow{3}{*}{\parbox[t]{1.5cm}{4\,890 {(367)} non-bursty papers}} &\multicolumn{2}{|l|}{ 62\% {(65\%)}  {exhibit usual ageing behaviour}}\\
&\multicolumn{2}{|l|}{ {18\%  (17.5\%)} are flashes-in-the-pan}\\
 &\multicolumn{2}{|l|}{ {20\% (17.5\%)} exhibit delayed activity}\\
\hline\hline
\multicolumn{3}{|c|}{\parbox[t]{8cm}{\textbf{Categorisation by overall attractiveness}}}\\
\hline
\multirow{3}{*}{\parbox[t]{1.5cm}{4\,890 {(367)} non-bursty papers}} &\multicolumn{2}{|l|}{{60\%} {(50\%)} {have typical overall attractiveness}}\\
\hhline{~--}
&\multirow{3}{*}{\parbox[t]{1.6cm}{40\%{(50\%)} are atypical}}%
&\parbox[t]{4.45cm}{{18\%} {(22\%)} are more attractive \vspace{0.8ex}}\\
&&\parbox[t]{4.45cm}{12\% {(15\%)} are less attractive \vspace{0.8ex}}\\
&&\parbox[t]{4.45cm}{10\% {(13\%)} change attractiveness \vspace{0.8ex}}\\
\hline
\end{tabular}}
\end{table}

{
Using download data from {\emph{EPL}}, for publications over a 6.5-year period, we developed a statistical picture for the attractiveness of the journal and its contents.
The resulting insights are complementary to those from other measures such as scientometrics, altmetrics and influmetrics and it is hoped that the work will inspire similar download analyses for other journals.
The synchronous aspect of downloads from the journal is well described by a
sum of two exponential decays corresponding to different types of download
activities (early and later in an average {\emph{EPL}} paper's life).
The model has predictive power for the long-term behaviour of paper downloads from the
journal.

{{We observed some difference in the ageing behaviour of downloads between non-OA and OA papers, mainly in terms of volumes of downloads and how fast downloads are accumulated.
Otherwise, the proportions of OA papers with different download patterns are very similar to those for non-OA papers. }}

A diachronous study shows that a very small proportion of {\emph{EPL}} papers are characterised by {high levels of} burstiness in download activity.
The ageing of the remaining non-bursty papers can be studied using download half-lives.
The term \emph{flash in the pan} applies to {about $18\%$} of publications and {a similar number} exhibit delayed recognition. The remaining papers have more usual ageing patterns.

The overall attractiveness of papers' entire lifespans can also be examined and, while the majority of papers accumulate downloads in a rather typical manner, 18\% {(22\% for OA papers)} have an especially high degree of attractiveness while 12\% {(15\% for OA papers)} are less attractive.
The attractiveness metrics for a smaller proportion of papers switches during their lifetimes.

One can speculate about the reasons for the various download {statistics} observed and  aspects such as novelty,  topicality and catchiness of title or a successful press release all can contribute to download characteristics.
Other factors include the target audience, whether the paper is specialized, broad or, indeed, on an unusual topic.
Similar to citations, different patterns of downloads are likely  for different disciplines.
Of course, one should be aware that not everything downloaded is actually read.
Nonetheless, as a first step on the road to scientific recognition, downloading is nowadays a widespread component of scientific activity, so its statistics contain information on the journal, its contents and, indeed, its   community of contributors and readers.
}

\section{Acknowledgements}
We thank {Daniel Barrett (IOP Publishing) and the staff of {\emph{EPL}}} for providing the data and assistance. This work was supported by the 7th FP, IRSES project No. 269139 ``Dynamics and cooperative phenomena in complex physical and biological environments'' and IRSES project No. 295302 ``Statistical physics in diverse realizations''.

\end{document}